\documentclass[prl,twocolumn,groupedaddress,showpacs]{revtex4}

\usepackage{graphicx} 

\newcommand{\figwidth}{0.90\columnwidth}

\begin{document}
\bibliographystyle{apsrev}

\title{Transport of {B}ose-{E}instein Condensates with Optical Tweezers}

\author{T.L. Gustavson}
\author{A.P. Chikkatur}
\author{A.E. Leanhardt}
\author{A. G{\"o}rlitz}
    \altaffiliation{Current address: Universit{\"a}t Stuttgart, 5. Phys. Inst., Germany}
\author{S. Gupta}
\author{D.E. Pritchard}
\author{W. Ketterle}
\homepage[Group website: ]{http://cua.mit.edu/ketterle_group/}
\affiliation{Department of Physics, MIT-Harvard Center for Ultracold
Atoms, and Research Laboratory of Electronics,
Massachusetts Institute of Technology, Cambridge, MA 02139}

\date{\today}

\begin{abstract}
We have transported gaseous Bose-Einstein condensates over distances
up to 44~cm.  This was accomplished by trapping the condensate in the
focus of an infrared laser and translating the location of the laser
focus with controlled acceleration.  Condensates of order 1~million
atoms were moved into an auxiliary chamber and loaded into a magnetic
trap formed by a Z-shaped wire.  This transport technique avoids the
optical and mechanical access constraints of conventional condensate
experiments and creates many new scientific opportunities.
\end{abstract}

\pacs{03.75.Fi, 32.80.Pj, 39.25.+k, 39.}

\maketitle


Since the achievement of Bose-Einstein condensation (BEC) in dilute
gases of alkali atoms in 1995, intensive experimental and theoretical
efforts have yielded a great deal of progress in understanding many
aspects of BEC~\cite{Ketterle1999a,Dalfovo1999a}.  Bose-Einstein
condensates are well controlled ensembles of atoms useful for studying
novel aspects of quantum optics, many-body physics, and superfluidity. 
Condensates are now used in scientific studies of increasing
complexity requiring multiple optical and magnetic fields as well as
proximity to surfaces.

Conventional condensate production techniques severely limit optical
and mechanical access for experiments due to the many laser beams and
magnetic coils needed to create BECs.  This conflict between cooling
infrastructure and accessibility to manipulate and study condensates
has been a major restriction to previous experiments.  So far, most
experiments are carried out within a few millimeters of where the
condensate was created.  What is highly desirable is a condensate
``beam line'' that delivers condensates to a variety of experimental
platforms.  Transport of charged particles and energetic neutral
particles between vacuum chambers is standard, whereas it is a
challenge to avoid excessive heating for ultracold atoms.  Thus far,
transport of large clouds of atoms has only been accomplished with
laser-cooled atoms at microkelvin
temperatures~\cite{Greiner2001,Kishimoto2001}.  Condensates are
typically a few orders of magnitude colder and hence much more
sensitive to heating during the transfer.

In this Letter, we demonstrate an application of optical tweezers that
can transfer Bose condensates over distances of at least 44~cm
(limited by the vacuum chamber) with a precision of a few micrometers. 
This separates the region of condensate production from that used for
scientific studies.  The ``science chamber'' has excellent optical and
mechanical access, and the vacuum requirements in this region may well
be less stringent than those necessary for production of BEC. This
technique is ideally suited to deliver condensates close to surfaces,
e.g. to microscopic waveguides and into electromagnetic cavities.  We
have used this technique to transfer condensates into a macroscopic
wiretrap~\cite{schm98quantumwire, fort98, dens99guide, dens99meso,
fort00, Key2000} located 36~cm away from the point where the
condensates were produced.

An alternative but less flexible method to create condensates close to
surfaces is to evaporatively cool directly in a wiretrap, as was
accomplished very recently~\cite{Ott2001,Reichel2001}.  Recently,
small condensates were also produced directly in an optical
trap~\cite{Barrett2001}, eliminating the complexities of magnetic
trapping.  Although evaporation in these small traps can be very fast
due to the tight confinement, the small trap volume fundamentally
limits the number of condensed atoms.  In contrast, the optical
tweezers method combines delivery of condensates into microtraps with
the well-established techniques of creating large condensates.

\begin{figure}
\includegraphics{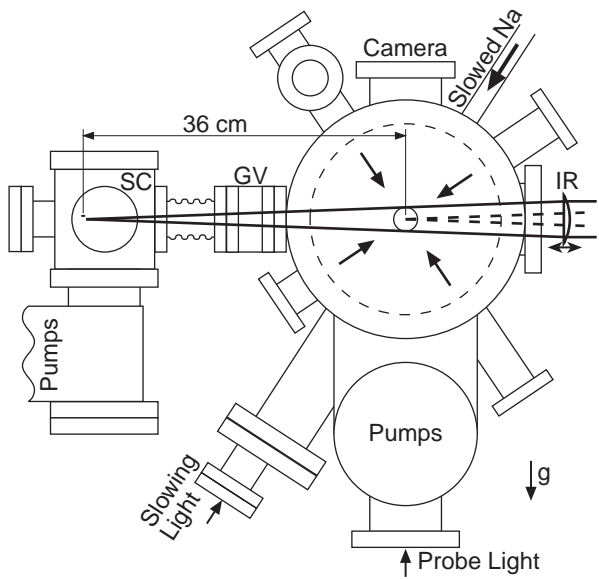}
\caption {Schematic of the apparatus, side view.  The science chamber
(labeled SC), is isolated by a gate valve (GV) from the trapping
chamber.  The large circle represents a 10" conflat nipple and the
dashed circle is the chamber wall.  The magnetic trap coils are
mounted in 6" recessed ports (not shown), bolted onto the 10" nipple. 
The central arrows represent 4 of the 6 orthogonal MOT beams.  The
other two MOT beams pass through the small windows indicated by the
small central circle.  The end of the Zeeman slower, shown at the top
right, is 33$^{\circ}$ from vertical.  The optical tweezers transfer
beam is horizontal, with its focus shifting from the MOT position a
distance of 36~cm to the Z-shaped wiretrap in the science chamber. 
Drawn to scale.}
\label{fig:schematic}
\end{figure}

The experiment was carried out in a new sodium condensation apparatus
that is an evolution of the original MIT design, which has been
described previously~\cite{Mewes1996a}.  The main challenge was to
integrate two additional viewports for the optical tweezers into the
stainless steel ultra-high vacuum (UHV) chamber.  The optical tweezers
beam had to be perpendicular to gravity, due to the relatively weak
axial confinement.  As a result, the atomic beam and the Zeeman slower
could not be arranged horizontally as in our previous BEC apparatus. 
The slower was placed at an angle of 33$^{\circ}$ from vertical.  The
magnetic trap coils were mounted outside the vacuum in recessed ports. 
A schematic is shown in Fig.~\ref{fig:schematic}.  The science chamber
was isolated from the trapping chamber by a gate valve and bellows. 
This allows the science chamber to be modified or even replaced
without compromising the UHV trapping chamber.

The trapping and science chambers were each pumped by separate ion and
titanium sublimation pumps, reaching pressures $\simeq 10^{-11}$~torr. 
The oven chamber was differentially pumped relative to the trapping
chamber and can support a factor of $10^{5}$ pressure difference.  The
Zeeman slower combined decreasing and increasing magnetic field
slowing (a so called ``spin-flip slower'') and delivered $> 10^{11}$
slowed atoms/s, and $\simeq 10^{10}$ atoms were loaded into a
dark-SPOT type MOT after 3~s~\cite{Ketterle1999a}.

Atoms were transferred into a Ioffe-Pritchard type magnetic trap
wound in a cloverleaf configuration with a radial gradient of
140~G/cm, axial curvature of 100~G/cm$^{2}$ and axial bias field of
100~G. The atoms were then compressed radially in 3.5~s by reducing
the axial bias field.  Forced RF evaporative cooling took a total of
30~s during which the gradient field was ramped from 140~G/cm to
280~G/cm in 4.5~s for additional compression, held constant at
280~G/cm for 21~s, then finally ramped down in 4.5~s to
140~G/cm to minimize three-body loss.  Typical
condensates contain 10 to 20 million atoms, and we expect to increase
these numbers with further optimization.
 
The condensate in the magnetic trap had to be decompressed
considerably in order to reduce density dependent losses during the
transfer and to improve spatial overlap with the optical tweezers
(optical dipole trap), because its long axis was along the radial
direction of the magnetic trap.  The magnetic trap was first
decompressed by increasing the axial bias field, which decreased the
radial trapping frequency from 200 to 85 Hz.  The condensate was
further decompressed by lowering the current by a factor of 10 in both
the gradient and curvature coils simultaneously.

The optical dipole trap was produced by focusing an infrared laser
(1064 nm) onto the center of the magnetic trap~\cite{stam98odt}.  The
output of the laser was spatially filtered by a single-mode fiber and
its intensity was adjusted with an acousto-optic modulator placed
before the fiber.  After the fiber, the beam was expanded and
collimated, and then focused by a 500~mm achromatic lens placed on a
translational stage.  This focus was imaged onto the condensate by a
relay telescope, yielding a $1/e^{2}$ beam waist radius at the
condensate of $w_{0}=24$~$\mu$m.

The condensate was transferred by ramping the infrared laser light
linearly up to 180~mW in 600~ms and then suddenly switching off the
decompressed magnetic trap.  The infrared beam was aligned
transversely to within $\sim 20$~$\mu$m of the condensate.  The
optical trap depth is proportional to $P/w_{0}^{2}$, where $P$ is the
power, and was 11~$\mu$K for 180~mW~\cite{counterrotate}.  The
transfer efficiency into the optical trap was close to 100\%.  The
laser light was then ramped down to 90~mW during the first second of
the transfer into the science chamber, in order to minimize three-body
loss.  The measured optical trap frequencies at 90~mW were 4~Hz
axially and 440~Hz radially.

The transport of the condensate to the science chamber was
accomplished by translating the 500~mm lens.  This was achieved using
a linear translation stage (MICOS/Phytron \#MT-150-400-DC220) with
400~mm maximum travel and 120~mm/s maximum velocity.  The motor was a
DC brushless servo type, with 0.5~$\mu$m encoder resolution.  A
feedback loop in the motor controller servoed the stage position to a
trajectory specified in terms of jerk (derivative of acceleration),
acceleration, velocity, and distance.  Efficient transfer requires
smooth, adiabatic motion.  We used a trapezoidal acceleration profile
that increased with constant jerk, had a flat top upon reaching
maximum acceleration, and decreased at constant jerk, followed by a
period of zero acceleration upon reaching maximum velocity.  For
deceleration, the opposite procedure was followed.  Initially, we
found that mechanical vibrations in the translation stage motion
caused loss of atoms due to severe heating.  This problem was
eliminated by adding two stages of vibration isolation using rubber
dampers and lead weights, which reduced the vibrations by a factor of
100.

Once accomplished, the transfer was quite robust and worked for a
range of motion parameters, up to 200~mm/s$^{2}$ acceleration, 80~mm/s
velocity, \mbox{$\sim 1000$~mm/s$^{3}$} jerk, and with transfer times as
short as 4~s.  The best transfer was achieved with the following
maximum values: jerk = 20~mm/s$^{3}$, acceleration = 37~mm/s$^{2}$,
and velocity = 70~mm/s, yielding a total transfer time of 7.5~s. 
Routinely we were able to transfer condensates with more than 6$\times
10^{5}$ atoms into the science chamber.  A complete systematic study
of motion parameters was not practical due to large shot-to-shot
fluctuations in the number of atoms transferred.  We attribute this to
alignment uncertainty of the optical tweezers due to variations in the
compression of the rubber vibration dampers when the stage was moved
back-and-forth.  We plan to eliminate this problem by installing a
smoother translation stage.

\begin{figure}
\includegraphics[width=\figwidth]{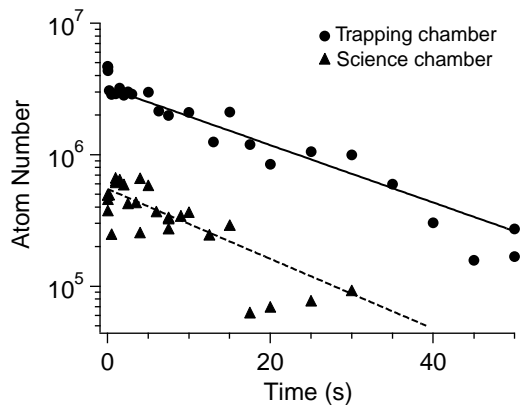}
\caption
{Lifetime of optically confined Bose-Einstein condensates in the
trapping and science chambers.  The number of condensed atoms is
plotted vs.\ trapping time.  Circles and triangles represent data in
the trapping and science chambers, respectively.  Both traps had the
same characteristics with 90~mW power and trap frequencies
of 4~Hz axially and 440~Hz radially.  The fluctuations in the science
chamber data are mainly due to alignment irreproducibilities in the
translation stage (see text).  The lines are exponential fits to the
data.  The lifetimes in the main and the science chambers were $20 \pm
2$ and $16 \pm 4$~s, respectively.}
\label{fig:odtlifetime}
\end{figure}

The lifetimes of the atoms in the optical dipole trap in both the 
trapping and science chambers are shown in Fig.~\ref{fig:odtlifetime}.
In the trapping chamber, the initial loss was due to three-body decay
and then the lifetime of $20 \pm 2$~s was limited mainly by background
gas collisions.  The noise in the lifetime data in the science chamber
was due to shot-to-shot fluctuations in the transfer.  The measured
lifetime of $16 \pm 4$~s was limited by background gas collisions. 
The number of transferred atoms was about four times lower than the
number remaining in the optical trap after simply holding atoms in the
trapping chamber for 7.5~s, which implies that the loss due to
transfer was comparable to the loss from the initial three-body decay. 
Therefore, a large volume dipole trap with an elliptical focus in
which three-body recombination is greatly diminished~\cite{1D2D}
should be able to deliver multi-million atom condensates.  First 
attempts to translate such an elliptical focus failed, probably due to 
lens aberrations.

\begin{figure}
\includegraphics{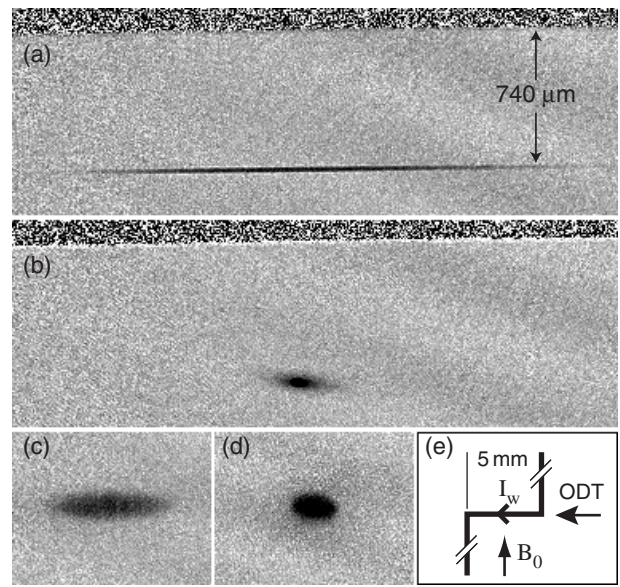}
\caption
{Absorption images of condensates in the science chamber, side
view.  All images have the same scale.  Condensates of $\simeq
6\times 10^{5}$ atoms in: a) optical trap and b) wiretrap.  The center
segment of the Z-shaped wire is visible as a dark speckled horizontal
strip and is 740~$\mu$m above the trapped atoms.  Condensate released
from: c) an optical trap after 10~msec time-of-flight and d) wiretrap
after 23~ms time-of-flight.  e) Schematic of the wiretrap, top view. 
$I_{w}=2$~A is the current through the wire, $B_{0}=2.9$~G is the bias
field.  Atoms are trapped below the 5~mm long central segment of the
wire, which is aligned with the optical trap axis.  The supporting end
segments, which provide field curvature, are truncated in the figure. 
The wiretrap was located 36~cm from where the condensates were
produced.}
\label{fig:traps}
\end{figure}

To demonstrate the utility of the optical tweezers transport, we
delivered condensates into a magnetic trap formed by a current $I_{w}$
in a Z-shaped macroscopic wire~\cite{reic99surface,haas01ztrap}
(diameter = 1.27~mm) and a bias field $B_{0}$ coplanar with the
Z-shaped wire and orthogonal to its central segment, as shown in
Fig.~\ref{fig:traps}.
The length of the central wire was $L=5$~mm, and the supporting end
segments were longer than 25~mm.  The trap position is located at
$z_{0}=(\mu_{0}/2\pi) I_{w}/B_{0}$ below the central wire, where the
external bias field is equal and opposite in direction to the magnetic
field produced by the wire.  The two end segments provide the axial
curvature $B''\propto B_{0}/L^{2}$ and produce a bias field at the
trap bottom $B_{bot}\propto
(I_{w}z_{0})/(4z_{0}^{2}+L^{2})$~\cite{haas01ztrap}.  The radial
gradient $B'$ is $(2\pi/\mu_{0})B_{0}^{2}/I_{w}$, and the radial trap
frequency is proportional to $B'^{2}/B_{bot}$.

The optical trap was aligned to overlap the original condensate in the
trapping chamber magnetic trap and to be about 1~mm below the wire in
the science chamber.  The current in the wiretrap and the current
producing the bias field $B_{0}$ were linearly ramped up in 1~s.  The
optical trap was then slowly ramped down to zero, transferring the
condensate into the magnetic wiretrap.  Nearly 100\% efficient
transfer was achieved for $B_{0}=2.9$~G and
$I_{w}=2.0$~A~\cite{bfields}.  The trap frequencies were measured to
be $36.0 \pm 0.8$~Hz radially and $10.8 \pm 0.1$~Hz axially.  The
lifetime of the condensate in the wiretrap was measured to be $5 \pm
1$~s.  This lifetime could probably be improved by adding a
radiofrequency shield to limit the trap depth~\cite{Mewes1996a}.  By
reducing the current in the wire, condensates were also moved to
within a few microns from the wire surface.

In conclusion, we have used an optical dipole trap as a tool for
moving and manipulating condensates over a range of 44~cm, and used
this tool to load atoms into a magnetic trap 36~cm away from the
center of the trapping chamber.  The ability to move and position
condensates allows them to be produced under optimal conditions, then
moved to another region to perform experiments requiring maximal
optical and mechanical access.  This flexibility will be important for
the next generation of BEC experiments, many of which may require
close proximity between condensates and other materials.

One key application that is hotly pursued by several groups is to load
condensates into micro-fabricated magnetic waveguides built using
lithographic wires on substrates, similar to integrated circuit
chips~\cite{wein95trap,reic99surface,mull99,dekk00,cass00nanofab,
folm00atomchip,Ott2001,Reichel2001}.  Such waveguides, analogous to
fiber optics for light, may lead to improved manipulation of
condensates, enabling sensitive atom interferometers.  Other
applications include studies of condensate-surface interactions
\cite{Shimizu2001,Henkel1999} and experiments which require extreme
magnetic shielding such as some proposed studies of spinor condensates
\cite{Law1998,Ho2000}.  Another possible application is producing a
continuous atom laser.  Previously demonstrated atom lasers work by
depleting a single condensate
\cite{Mewes1997a,Anderson1998a,Bloch1999a,Hagley1999a}.  A continuous
atom laser could be produced by repeatedly transferring condensates
into a reservoir from which atoms are continually out-coupled.  The
dipole trap could be used to transfer atoms into an optical or
magnetic trap reservoir that is spatially separated from the
condensate production region to avoid losses due to scattered light. 
Finally, one could move condensates into high finesse optical or
microwave cavities.

\begin{acknowledgments}
This work was funded by ONR, NSF, ARO, NASA, and the David and Lucile
Packard Foundation.  A.E.L., A.P.C., and A.G.\ acknowledge additional
support from NSF, JSEP, and DAAD, respectively.  We are indebted to
T.~Rosenband and S.~Inouye for their important contributions to
the apparatus.  We thank D.~Schneble for a critical reading of the 
manuscript.
\end{acknowledgments}


\begin{thebibliography}{33}
\bibitem[{bfi()}]{bfields}
\bibinfo{note}{The residual magnetic fields in the science chamber, estimated
  to be about 300 mG, were not compensated.}
\end{thebibliography}

\end{document}